\documentclass{ws-ijmpcs}

\begin{document}


\markboth{Goran S. Djordjevic} {Noncommutativity and Humanity -–
Julius Wess and his Legacy}

%
\catchline{}{}{}{}{}
%

\title{NONCOMMUTATIVITY AND HUMANITY –- JULIUS WESS AND
HIS LEGACY}

\author{GORAN S. DJORDJEVIC}

\address{Department of Physics, Faculty of Science and Mathematics, University of Ni\v s\\
 P.O. Box 224, 18000 Ni\v s, Serbia\\
gorandj@junis.ni.ac.rs}

\maketitle

\begin{history}
\received{10 March 2012} 
\end{history}

\begin{abstract}
A personal view on Julius Wess's human and scientific legacy in
Serbia and the Balkan region is given.
 Motivation for using noncommutative and
nonarchimedean geometry on very short distances is presented. In
addition to some mathematical preliminaries, we present a short
introduction in adelic quantum mechanics in a way suitable for its
noncommutative generalization. We also review the basic ideas and
tools embedded in $q$-deformed and noncommutative quantum mechanics.
A rather fundamental approach, called deformation quantization, is
noted. A few relations between noncommutativity and nonarchimedean
spaces, as well as similarities between corresponding quantum
theories, in particular, quantum cosmology are pointed out. An
extended Moyal product in a frame of an adelic noncommutative
quantum mechanics is also considered.

\end{abstract}

\ccode{PACS numbers: 01.60.+q, 01.78.+p, 02.40.Gh, 98.80.-k,
98.80.Qc}

\section{Julius Wess - Scientist and Humanist}

The role of Julius Wess in the renewal and development of research
in the field of high energy physics, especially in Serbia and
Croatia, but also in other parts of former Yugoslavia and beyond,
and in particular in what we shall refer to as international
cooperation, is significant and large. This positive influence can
be felt even now and will continue although he has passed away.
The significance of his human and scientific mission in this part
of the world is not well enough known in many respects. That is
even in Germany and in the institutions in which he worked and
which he managed in those years, and from which he drew the
logistic, financial and every other form of support for this
mission, above all the Max Planck Institute (MPI) for Physics and
Ludwig Maximilian University (LMU) Munich. In one segment of this
paper I shall attempt, from my own point of view and on the basis
of my knowledge and memories, to shed light on the period of our
friendship and cooperation in the period from 2000 to 2007. It is
my desire that this aspect of his noble personality and immaterial
legacy be preserved for remembrance as a valuable example for the
future, as well as to provide an overview of the processes which
were initiated thanks to Julius Wess and which continue to this
day.

\subsection{Initial information and contacts}

Much can be learned about the process of the forming of the idea
and beginnings of Julius' initiative Scientists in Global
Responsibility (Wissenschaftler in Global Verantwortung –- WIGV in
German), from a valuable text in this volume by Lutz
M\"{o}ller,\cite{moeller} who was a student of Julius', and also a
very important associate in the period from 1999 to 2003.

The first information about the WIGV initiative and two
postdoctoral scholarships in Munich in the field of theoretical
physics and mathematics for candidates from former Yugoslavia
reached me early in March of 2000 through Branko Dragovi\' c (my
supervisor in the period from 1989 to 1999), thanks to his
teaching engagement at the University of Banja Luka and the
information he received there.  By the time I had made first
contact and applied (with the recommendation of B. Dragovi\' c),
which went via Martin Schottenloher, Julius' colleague from the
Department of Mathematics at LMU, both scholarships had been
awarded to two mathematicians. Julius' desire to support as many
young people as possible, his influence and ability to quickly
find the most favorable solution, led to a sequence of events
which, following my first lecture given early in July in Munich,
resulted in my receiving the position of guest researcher, which
lasted, with short breaks, almost two years, until the end of
November 2002.

\subsection{WIGV and scholarships}

Scholarships for students and researchers from former Yugoslavia
were probably one of the most significant segments of Julius'
contribution to the improvement of research in the region, above all
in the field of theoretical physics, as well as mathematics.

As early as 1999, Igor Bakovi\' c from the Faculty of Mathematics
and Institute Rudjer Bo\v skovi\' c in Zagreb received a full
scholarship for a doctorate in the field of applications of
noncommutative geometry on short distance physics. As has already
been mentioned, in the course of 2000, as part of a special DFG
program (Noncommutative fields in physics) two postdoctoral
scholarships were given to Svjetlana Terzi\' c from Podgorica and
Bo\v zidar Jovanovi\' c from Belgrade. In addition to myself,
somewhat later, and funded from other sources, mobility grants to
Munich were awarded to Goran Duplan\v ci\' c, Josip Trampeti\' c,
Dragutin Svrtan and Larisa Jonke from Zagreb, Marija Dimitrijevi\'
c, Maja Buri\' c and Voja Radovanovi\' c from Belgrade. In those
years, numerous researchers stayed in Munich from Montenegro,
Bulgaria, Ukraine, Russia, Morocco etc.

Almost all scholarships were realized via the LMU and/or MPI for
Physics in Munich. Julius thus made it possible, especially but not
exclusively, for younger researchers to spend significant time
working and improving their research capacities under very favorable
circumstances in his group.

A range of other programs and conferences which he led or
organized (especially workshops in Bayrischzell) which were
continued even after his death, made possible meetings and
cooperation between numerous researchers from the region. Maybe
most significantly, he made possible to us to establish contacts
and cooperation with many excellent researchers from all around
the world, who were in various ways linked to Julius and who
visited Munich.

In addition to the positive effect on the scientific orientation
and therefore the careers of the researchers named and those
unnamed, as complete and precise documentation is not available,
these programs had a range of indirect, sometimes no less
significant effects. Knowledge was transferred not only to
individuals but to the institutions where researchers were
employed. it made it possible for their students and colleagues to
be a part research international programs and to cooperate with
many other institutions in Germany, Europe and around the world.

Likewise, new contacts were established and some others
reinvigorated between individuals and scientific centers in the
Balkans region. We emphasize this region as the text deals mainly
with the influence which Julius in the last count achieved in the
Balkans, that is, in Southeast Europe (SEE), although the term SEE
is geographically not entirely correct.

\subsection{The organization of conferences in the region}

It must be emphasized that in Serbia, focussing for a moment only on
this republic of the former Yugoslavia, for at least one whole
decade, from 1991 to 2001, not a single international conference had
been held, no school, workshop or any other form of scientific
meeting, in the field of theoretical physics and mathematical
physics on high energies. Researchers and students in Serbia were
practically without the possibility of participating in similar
meetings abroad, not to mention in the West. Therefore, Julius'
idea, launched in October 2000 for organizing the first
German-Serbian school of modern mathematical physics was of great
importance. The summer school was held in Sokobanja, Serbia (in
August of 2001), and it was much more than yet another Summer School
somewhere in Europe.

The School was organized and took place an effort that was truly
pioneering and with a great desire that after a horrible decade in
former Yugoslavia, in all areas of life as well as in physics. The
aim was to make it possible for students, above all, to attend an
international school of mathematical physics of the highest level,
with the partici\-pa\-tion of a larger number of lecturers and
students from different countries. Main financial support came
only two days before the beginning of the School, thanks to
Julius' efforts and standing which he had with the
Bundesministerium f\"ur Bildung and Forschung (Federal Ministry of
Education and Research -- Germany). The school had 65
participants, 13 lecturers had cycles of three to four lectures,
while another 13 had lectures which were closer in form to
workshops.

Although Julius was unable to attend the School, seven students from
Germany participated in it, above all his PhD students from Munich.
For me, as someone who had grown up in a different culture, it took
time and trust to believe they had come to Serbia so soon after all
the wars, led only by own free will, attracted either by the program
offered and recommended by their supervisor. This was a pleasant and
valuable discovery. In addition to a varied program of science and
teaching, there was a social program, and the socializing had an
irreplaceable positive effect on later closeness, trust and
cooperation between colleagues from different countries.

The organization and success of the School, in addition to the
great roles of Julius Wess and Lutz M\"{o}ller were in large part
also due to the efforts of the Institute for Physics in Belgrade,
above all Branko Dragovi\' c and Dragan Popovi\' c, as well as my
colleagues and associates from the Department of Physics of the
University of Ni\v s, Ljubi\v sa Ne\v si\' c, Jelena Stankovi\' c
and Dragoljub Dimitrijevi\' c.

Although at some moments in the preparation it seemed as if the
organization of the entire 12-day School hung by a thread, and
communication among organizers from different countries, cultures
and mentalities, who were practically cooperating for the first time
in this way, was put to the test, the impression that the School was
very successful was confirmed only a few days after its completion
at the very next 8th Adriatic meeting held in Dubrovnik, Croatia.

The story of the Adriatic meeting is beyond the immediate
framework of the subject, however, it should be mentioned that the
Adriatic meeting in the 1970s and 1980s was an international
scientific conference of the highest level which was so
significant that it was clearly noted beyond the boundaries of
former Yugoslavia. I believe that nostalgia for the good old times
and conferences which Julius himself had attended on several
occasions at a time when tensions between the different peoples of
former Yugoslavia were not felt and noticed as much, was a motive
for Julius to place the revival of the Adriatic meeting, in
addition to other programs, high on the list of priorities of the
WIGV initiative. In brief, Adriatic meeting 2001 was an excellent,
valuable conference, excellently organized by colleagues from
Croatia led by Josip Trampeti\' c.

It is significant for our topic that in conversations with Julius
and colleagues from Croatia (J. Trampeti\' c, N. Bili\' c, B.
Guberina and others) it was agreed that the second German-Serbian
school of modern mathematical physics should be held again in 2002
in Serbia (not in 2003 as previously agreed), so that the School
(focussed on mathematical physics) and Adriatic meeting (with an
accent on phenomenology in high energy physics) could be organized
in even or uneven years respectively, and in this way could address
the need of both the training of young researchers and exchange of
knowledge and achievements among leading researchers.

A situation in which Serbian and Croatian physicists talk with an
Austrian-German physicist, support one another, find individual
and shared interests which are almost identical, was for me, as
someone new to these types of activities, something of a novelty
and very encouraging. This was probably the decisive influence for
the idea of forming some kind of network (which would later
receive its acronym SEENET-MTP). It should also be mentioned that
the later development of events would show that we were all a bit
too optimistic about the range and speed of revival of scientific
life in our region.

Early in 2002, within the inner circle of physicists in Serbia there
came about a disagreement about the organization of the Second
School and the need for and ways of forming a regional network in
Mathematical and Theoretical Physics. In the next four to five years
this would lead to a separation in the organization of the 2nd and
3rd and partially the 4th School of Mathematical Physics on the one
hand, and the founding of the SEENET-MTP network and its meetings,
above all BW2003, BW2005, BW2007 etc. Julius, in accordance with his
health, which took a dramatic turn for the worse in 2003, and
opportunities for influence after retiring, assisted all the above
meetings and activities greatly. He also supported a few meetings in
pure Mathematics in the region.

\subsection{Forming of the SEENET-MTP  Network}

This text is not a text on the SEENET-MTP network, but on Julius'
contribution to the revival of scientific life and cooperation in
some fields of science in the Balkans, hence only some basic data
will suffice.

Staying and working in Munich, in Julius' group, at the LMU and
MPI was a fantastic experience. Although I already had some
experience in a rich program of seminars and lectures given by
great scientists at the Mathematics Institute Steklov in Moscow, a
new quality appeared here. Guest lecturers were most often
connected to corresponding groups in Munich through international
projects and networks of institutions and research groups. Julius'
wish was to help, as much as possible, in connecting scientists
from former Yugoslavia. My impression was that earlier, and in
particular in 2002, a "critical mass" of students and researchers
in this region could not have been achieved and that a larger
context should be attempted - the Balkans, which would include
Turkey, Greece, Bulgaria, Romania etc. In my belief this kind of
approach would in addition to a scientific also have a political
dimension and other numerous benefits. Agreement was quickly
reached and the name Southeastern European Network in Mathematical
and Theoretical Physics (SEENET-MTP, www.seenet-mtp.info) was
created in an informal atmosphere ... in the triangle
Wess-Trampeti\' c-Djordjevi\' c, in the vicinity of Alta and Neue
Pinakotek, in Munich, sometime early in July of 2002. With Julius'
personal recommendations in this period I visited CERN, ICTP,
UNESCO headquarters in Paris and the UNESCO Office for Europe in
Venice. In the course of these visits and talks the foundations
were laid for support for the future SEENET-MTP network. Let
mention just a few the most supportive persons from UNESCO, Dr.
Vladimir Kouzminov, Dr. Maciej Nalecz and Dr. Vladimir Zharov.

The founding meeting of the network was envisaged as a workshop
Balkan Workshop - BW2003, devoted to Mathematical, Theoretical and
Phenomenological Challenges Beyond the Standard Model, with the
addition of Perspectives of Balkans Collaboration and as a
satellite meeting the Fifth General Conference of the Balkan
Physical Union (Vrnja\v cka Banja, Serbia, August 2003). This made
it possible for the meeting to have a regional dimension in many
respects, to have present representatives of practically all
countries from the region and to secure savings, in particular in
travel expenses. Unlike the First School and some other actions
the contribution to the budget of BW2003 from Germany was not more
than one third. Significant contributions came from DFG (with
Julius' support) and DAAD (with the support of Dieter L\"{u}st).
It was for us a very difficult but important schooling in the
writing of several applications, projects and their
implementation. This scientific meeting with excellent lecturers
 was completed with the ratification of a letter of intent,\cite{intention}
 my election as Coordinator of the Network and Julius' election
 as Coordinator of the Network`s Scientific Advisory Committee (SAC).

While singling out the role of an individual might seem
disproportionate, the role of Boyka Aneva in motivating colleagues
and institutions from Sofia and Mihai Visinescu for those from
Romania was crucially important.

The publishing of the Proceedings of BW2003, which consisted of
almost all invited lectures, was a formidable task for us. Without
Julius' authority this endeavor would not have been successful.
With the joint efforts of the editors, authors and volunteers from
Ni\v s, the Proceedings\cite{procbw2003} was published in
cooperation with World Scientific and found its place in libraries
around the world and was the scientific ticket and identity card
of the SEENET-MTP network and its scientific potential in the next
few years.

Less than two months after BW2003 Julius had a serous
cardiovascular accident in Vienna and our first meeting after that
was in Munich in January 2004. We spent more time discussing my
fight with stress than his condition, operation and almost
miraculous recovery in a relatively short period of time! It has
been a great pleasure to mention that early in November of 2003,
by the proposal of several physicists from Serbia, Julius was
elected a member of the Serbian Academy of Science and Arts.

From a range of joint activities over the next years, I would like
to single out his participation at the, small in terms of number of
participants but significant, workshop Quantum Models - QM2005,
which was held in Ni\v s in November 2005. At this meeting the
structure of the SEENET-MTP network was laid out and its Terms of
References were ratified as the main regulatory act for the
functioning of the network until the present day.

All scientific meetings in the period from 2003 to 2007 were
organized with Julius' help, despite his health problems, about
which he almost never complained. All scientific publications,
above all proceedings, were prepared and published under his
expert guidance. Several projects which we realized with UNESCO
were also joint efforts with Julius. Adding all exchange programs
in this period, of course without counting conferences, around 50
stays of scientists at foreign institutions in South East Europe
could be realized in the framework of SEENET-MTP, as well as a
number of outstanding guest lecturers from Western Europe and the
USA

With the ending of the Balkan Summer Institute BSI2011, one of the
four segments of which was a Workshop devoted to the Scientific
and Human Legacy of Julius Wess, SEENET-MTP has from the outset 50
individual founders and three institutions (Bucharest, Sofia, Ni\v
s) grown to a network of 19 institutions in ten countries of the
Balkans (only Montenegro is missing), with around 300  individual
members and 12 partner institutions from around the world. With
some ten projects realized with support from UNESCO, ICTP,
Bavarian Ministry for Science and other institutions, and a
similar number of scientific meetings, now not only in Serbia and
Croatia, but also in Bulgaria, Romania, Turkey, and several joint
monographs and proceedings, as well as many individual and joint
papers referring to the SEENET-MTP (and related projects), despite
all the weakness and problems, we can say that we have much more
than we had in 2000. A large gap has been left with Julius'
passing but has been somewhat made up by the activities of the new
coordinator of SAC Goran Senjanovi\' c, the president of the
Network and its Representative Committee as of 2009, Radu
Constantinescu and Michael Haack from LMU Munich in many ways.
Some one hundred students who have passed through these programs
and activities, we hope, will go on to raise the level of
scientific research in the region to a higher one than our
generation has been able to achieve. In every one of these
successes will be a part of Julius' idea from 1999 and a part of
his personal benevolence.

\subsection{In place of a conclusion}

Julius Wess has in many respects influenced scientific life and the
work of a significant number of physicists and mathematicians from
former Yugoslavia in changing and improving it at a moment when it
seemed that there was no help coming and when our societies had
fallen into some forms of poverty worse than just material. In
addition to direct financial support (SINYu/SINSEE - modern intranet
for the Faculties of Science in Belgrade, Ni\v s and Novi Sad, and
access to high speed internet as an investment of around one million
Euros, the greatest donation in books for physics in Ni\v s ever, in
addition to similar donations in Tuzla and Banja Luka\cite{moeller},
great moral motivation to begin and persevere with a unique project
of the High school class for students with special abilities in
physics in Ni\v s,\cite{odeljenje} his suggestions and proposals
have opened the doors of many institutions and foundations for
individual as well as strategic programs.

Perhaps most importantly, he transferred to us his view of physics,
his way of thinking and doing research, encouraged us to address
serious problems, to approach great scientists and with his personal
example, taught us to be humble with and approachable for those
younger and not as well known.

In the end, a special personal note, his warm attitude, and the
attitude of his wife Traudi, towards my family and in particular
the children is something I will always remember.

\section{Noncommutative vs Nonarchimedean Geometry at Short
Distances - Motivation}

It is widely accepted that the standard picture of space-time should
be changed around and beyond Planck scale. ``Measuring'' of
spacetime geometry under distances smaller than Planck length $l_p$
is not accessible even to Gedanken experiments. It serves the idea
of ``quantization'' and "discretization" of spacetime and a natural
cutoff when using a quantum field theory to describe related
phenomena. We are pointing out two possibilities for a reasonable
mathematical background of a quantum theory on very small distances.
The {\it first} one comes from the idea of spacetime coordinates as
noncommuting operators\cite{djo-wess}
\begin{equation}
\label{djo-noncom}
[\hat x^{i},\hat x^{j}]=i\theta^{ij}.
\end{equation}
Some noncommutativity of configuration space should not be a surprise in
physics since quantum phase space with the canonical commutation relation
\begin{equation}
\label{djo-com}
[\hat x^i,\hat k^j]=i\hbar\delta^{ij} ,
\end{equation}
where $x^i$ are coordinates and $k^j$ are the corresponding momenta
is the well-known example of noncommutative (pointless) geometry.
This relation is connected in a natural way with the uncertainty
principle and a ``fuzzy'' spacetime picture at distances
$~\theta^{1/2}$. Although it seems to make good physical sense for
$\theta^{1/2}\sim l_p$, characteristic noncommutative distances
could be related to gauge couplings,\cite{djo-wess} closer to
observable distances. It should be noted that the deriving of
uncertainty relation ($\delta x>l_p$) leads to a ``strange'' notion
of quantum line and probably beyond archimedean geometry, because a
coordinate always commutes with itself!\cite{djo-av}

The {\it second} promising approach to physics at the Planck scale,
based on non-archimedean geometry, was suggested.\cite{djo-volo} The
simplest way to describe such a geometry (often referred to as
ultrametric) is by using $p$-adic number fields $\Bbb Q_p$ ($p$ is a
prime). On the basis of the Ostrowski theorem\cite{djo-vladi} there
are no other nontrivial possibilities (besides the field of real
numbers $\Bbb R$) to complete the field of rational numbers $\Bbb Q$
with respect to a (nontrivial) norm.

There have been many interesting applications of $p$-adic numbers
and non-Archimedean geometry in various parts of modern theoretical
and mathematical physics (for a review, see
Refs.~\refcite{djo-vladi,djo-breke,DKhKV}). However, we restrict
ourselves here mainly to $p$-adic\cite{djo-vladi2} and
adelic\cite{djo-mentor} quantum mechanics (QM). It should be noted
that adelic QM have appeared quite useful in quantum
cosmology.\cite{vol2,58bw2007} The appearance of space-time
discreteness in adelic formalism (see, e.g. Ref.~\refcite{vol2}), as
well as in noncommutative QM, is an encouragement for further
investigations. The $p$-adic analysis and noncommutativity also play
a role in some areas of "macroscopic" physics.\cite{djo-jackiw}

The interplay between noncommutative and nonarchimedean approach to
quantum theory on very short distances was the main topic of my
research during "Munich period" and subject of numerous discussions
with Julius and his collaborators. This investigation continues to
be among my main research topics till now.

This paper is organized as follows: a personal view on Julius Wess
human and scientific legacy in Serbia and Balkan region is followed
by this note about motivation for investigation on a possible role
of noncummatative and nonarchimedean geometry in physics. A short
introduction to $p$-adic numbers, adeles and their functions is
given in section 3. This section is completed by a review of
$p$-adic and adelic QM based on the Weyl quantization and Feynman's
path integral. Previously observed and originally discovered
relations between noncommutative and $p$-adic QM are discussed in
section 4. A particular result on a new observed relation between an
ordering on commutative ring in frame of deformation
quantization\cite{djo-waldmann,64petrovac} and an ordering on
$p$-adic spaces with the intention to develop path integration on
$p$-adics by "slicening" of trajectories\cite{djo-zele} is
emphasized. Section 5 is devoted to a proposal for formulation of
noncommutative adelic QM and some aspects of the extended Moyal
product.\cite{kijev} This paper closes with a presentation of an
application of commutative and noncommutative mechanics in cosmology
on real and $p$-adic spaces with notes on some ideas for further
investigation, in particular, towards tachyonic cosmological
(inflatory) models.\cite{fdp}

\section{$p$-Adic Numbers, Adeles and Quantum Mechanics}

Any $p$-adic number $x\in \Bbb Q_p$ can be presented in the
form\cite{djo-vladi}
\begin{equation}
\label{djo-canonexp}
x = p^\nu(x_0+x_1p+x_2p^2+\cdots)\ ,\quad \nu\in \Bbb Z,
\end{equation}
where $x_i = 0,1,\cdots,p-1$ are digits and $p$ is a prime number.
$p$-Adic norm of any term $x_ip^{\nu+i}$ in the canonical expansion
(\ref{djo-canonexp}) is $|x_ip^{\nu+i}|_p =p^{-(\nu+i)}$ and the
strong triangle inequality holds,
 {\it i.e.} $|a+b|_p\leq\hbox{max}\{|a|_p,|b|_p\}$. It follows that $|x|_p = p^{-\nu}$ if
$x_0\neq 0$. Derivatives of $p$-adic valued functions $\varphi :
\Bbb Q_p\to \Bbb Q_p$ are defined as in the real case, but with
respect to the $p$-adic norm. There is no integral
$\int\varphi(x)dx$ in a sense of the Lebesgue
measure,\cite{djo-vladi} but one can introduce $\int^b_a\varphi(x)dx
= \Phi(b)-\Phi(a)$ as a functional of analytic functions
$\varphi(x)$, where $\Phi(x)$ is an antiderivative of $\varphi(x)$.
In the case of map $f:\Bbb Q_p\to \Bbb C$ there is well-defined Haar
measure. One can use the Gauss integral
\begin{equation}
\int_{\Bbb Q_\upsilon}\chi_\upsilon(ax^2+bx)dx =
\lambda_\upsilon(a)|2a|^{-{1\over2}}_\upsilon{\chi}_v
\big(-{b^2\over4a}\big)\ ,\quad a\not=0,\ \upsilon =\infty,2,3,5,\cdots\!,
\end{equation}
where index $\upsilon$ denotes real ($\upsilon =\infty$) and
$p$-adic cases, $\chi_\upsilon$ is an additive character:
$\chi_\infty(x)=\exp(-2\pi ix)$, $\chi_p(x) = \exp(2\pi
i\{x\}_p)$,\quad  where $\{x\}_p$ is the fractional part of $x\in
\Bbb Q_p$, and $\lambda_\upsilon (a)$ is the complex-valued
arithmetic function.\cite{djo-vladi} An adele\cite{djo-geljfand} is
an infinite sequence $a=(a_\infty, a_2,..., a_p,...)$, where
$a_\infty\in \Bbb R\equiv \Bbb Q_\infty$, $a_{p}\in \Bbb Q_{p}$ with
a restriction that $a_{p} \in \Bbb Z_{p}$ for all but a finite set
$S$ of primes $p$. The set of all adeles $\Bbb A$  may be  regarded
as a subset of direct topological product $\Bbb
Q_\infty\times\prod_p \Bbb Q_p$. $\Bbb A$ is a topological space,
and can be considered as a ring with respect to the componentwise
addition and multiplication. An elementary function on adelic ring
$\Bbb A$ is
\begin{equation}
\label{djo-elementary}
        \varphi (x)=\varphi _{\infty}(x_{\infty})\prod_{p}^{}\varphi
_{p}(x_{p})
        =\prod_{v}^{} \varphi _{v}(x_{v})  \;\;,
\end{equation}
with the main restriction that
$\varphi (x)$ must satisfy $\varphi_{p}(x_{p})=\Omega
(|x_{p}|_{p})$
for all but a finite number of $p$.
Characteristic function on
$p$-adic integers $\Bbb Z_p$
is defined by $\Omega (|x|_p) =1, \ 0\leq |x|_p \leq 1$ and
$\Omega (|x|_p) =0, \ |x|_p>1$.

The Fourier transform of the characteristic
function (vacuum state) $\Omega(|x_p|)$ is $\Omega(|k_p|)$.
All finite linear combinations of elementary functions
(\ref{djo-elementary}) make  the set $\cal D(\Bbb A)$ of the Schwartz-Bruhat
functions. The Hilbert space $L_2(\Bbb A)$ is a space of complex-valued functions
$\psi_1(x)$, $\psi_2(x),\dots$, with the scalar product and norm.


In foundations of standard QM one usually starts with a
representation of the canonical commutation relation
(\ref{djo-com}). In formulation of $p$-adic QM\cite{djo-vladi2} the
multiplication $\hat q\psi\rightarrow x \psi$ has no meaning for
$x\in\Bbb Q_p$ and $\psi(x)\in \Bbb C$. In the real case momentum
and hamiltonian are infinitesimal generators of space and time
translations, but, since $\Bbb Q_{p}$ is totally disconnected field,
these infinitesimal transformations become meaningless. However,
finite transformations remain meaningful and the corresponding Weyl
and evolution operators are $p$-adically well defined.

Canonical commutation relation (\ref{djo-com}) in $p$-adic case can
be represented by the Weyl operators ($h=1$)
\begin{equation}
\hat Q_p(\alpha) \psi_p(x)=\chi_p(\alpha x)\psi_p(x),
\end{equation}
\begin{equation}
\hat K_p(\beta)\psi(x)=\psi_p(x+\beta),
\end{equation}
\begin{equation}
\hat Q_p(\alpha)\hat K_p(\beta)=\chi_p(\alpha\beta)\hat K_p(\beta)\hat Q_p(\alpha).
\end{equation}
It is possible to introduce the family of unitary operators
\begin{equation}
\hat W_p(z)=\chi_p(-\frac 1 2 qk)\hat K_p(\beta)\hat Q_p(\alpha), \quad
z\in\Bbb Q_p\times\Bbb Q_p,
\end{equation}
that is a unitary representation of the Heisenberg-Weyl group.

The dynamics of a $p$-adic quantum model is described by a unitary
operator of evolution $U(t)$ formulated in terms of its kernel
${\cal K}_t(x,y)$. In this way\cite{djo-vladi2} $p$-adic QM is given
by a triple $(L_2(\Bbb Q_p), W_p(z_p), U_p(t_p))$.

Keeping in mind that standard QM can be also given as the
corresponding triple, ordinary and $p$-adic QM can be unified in the
form of adelic QM\cite{djo-mentor}
\begin{equation}
(L_2(A), W(z), U(t)),
\end{equation}
$L_{2}(\Bbb A)$ is the Hilbert space on $\Bbb A$, $W(z)$ is a unitary
representation of the Heisenberg-Weyl group on $L_2(\Bbb A)$ and
$U(t)$ is a  unitary representation of the
evolution operator on $L_2(\Bbb A)$.
The evolution operator $U(t)$ is defined by
\begin{equation}
U(t)\psi(x)=\int_{\Bbb A} {\cal
K}_t(x,y)\psi(y)dy=\prod\limits_{v}{} \int_{\Bbb Q_{v}}{\cal
K}_{t}^{(v)}(x_{v},y_{v})\psi^{(v)}(y_v) dy_{v}.
\end{equation}
Note that any adelic eigenfunction  has the form
\begin{equation}
\label{djo-psi}
\Psi(x) =
\Psi_\infty(x_\infty)\prod_{p\in S}\Psi_p(x_p)
\prod_{p\not\in S}\Omega(|x_p|_p) , \quad x\in \Bbb A,
\end{equation}
where $\Psi_{\infty}\in L_2(\Bbb R)$,
$\Psi_{p}\in L_2(\Bbb Q_p)$.
In the low-energy limit adelic
QM becomes ordinary one.

A suitable way to calculate propagator in $p$-adic QM is by $p$-adic
generalization of Feynman's path integral.\cite{djo-vladi2} There is
no natural ordering on $\Bbb Q_p$. However, a bijective continuous
map $\varphi$ from the set of $p$-adic numbers $\Bbb Q_p$ to the
subset $\varphi(\Bbb Q_p)$ of real numbers $\Bbb R$ was
proposed.\cite{djo-vladi} This map can be defined by (for an older
injective version see Ref.~\refcite{djo-zele})
\begin{equation}
\label{map}
\varphi(x)=|x|_p \sum_{k=0}^\infty x_k p^{-2k}.
\end{equation}
Then, a linear order on $\Bbb Q_p$ is given by the following
definition: $x<y$ if $|x|_p<|y|_p$ or when $|x|_p =|y|_p$ there
exists such index $m\geq0$ that digits satisfy $x_0 = y_0, x_1 =
y_1, \cdots,x_{m-1} = y_{m-1}\ ,x_m<y_m$. One can say: $\varphi(x) >
\varphi(y)$ iff $x>y$.

In the case of harmonic oscillator,\cite{djo-zele} it was shown that
there exists the limit
\begin{eqnarray}
{\cal K}_p(x^{\prime\prime},t^{\prime\prime};x^\prime,t^\prime)=
\lim_{n\to\infty} {\cal
K}_p^{(n)}(x^{\prime\prime},t^{\prime\prime};x^\prime,t^\prime) =
\lim_{n\to\infty}N^{(n)}_p(t^{\prime\prime},t^\prime) \nonumber \\
\times\int_{\Bbb Q_{p}}\cdots
\int_{{\Bbb Q}_{p}}
\chi_p\bigg(-{1\over h}\sum^n_{i=1}\bar S(
q_i,t_i;q_{i-1},t_{i-1})\bigg)dq_1\cdots dq_{n-1}\ ,
\end{eqnarray}
where $N^{(n)}_p(t^{\prime\prime},t^\prime)$ is the corresponding
normalization factor for the harmonic oscillator. The subdivision of
$p$-adic time segment $t_0<t_1<\cdots<t_{n-1}<t_n$ is made according
to linear order on $\Bbb Q_p$. In a similar way we have calculated
path integrals for a few quantum models. Moreover, we were able to
obtain general expression for the propagator of the systems with
quadratic action (for the details see Ref.~\refcite{djo-general}),
without ordering
\begin{equation}
\label{djo-quadratic} {\cal
K}_p(x^{\prime\prime}\!,t^{\prime\prime}\!;x^\prime\!,\!t^\prime)=
\!\lambda_p \!\bigg(\!-\frac{1}{2h}{\partial^2\bar S\over \partial
x^{\prime\prime}\partial x^\prime} \bigg)\!
\Big\arrowvert\frac{1}{h}{\partial^2\bar S\over \partial
x^{\prime\prime}\partial x^\prime} \Big\arrowvert^{\frac{1}{2}}_p
\chi_p \!\bigg(\!-{1\over h}\bar
S(x^{\prime\prime}\!,t^{\prime\prime}\!;x^\prime\!,\!t^\prime)
\bigg)\!.
\end{equation}
Replacing an index $p$ with $v$ in (\ref{djo-quadratic}) we can
write quantum-mechanical amplitude ${\cal K}$ in ordinary and all
$p$-adic cases in the same, compact (and adelic) form.

\section{Relations Between Noncommutative and $p$-Adic QM}

Noncommutative geometry is geometry which is described by an
associative algebra ${\cal A}$, which is usually noncommutative and
in which the set of points, if it exists at all, is relegated to a
secondary role. Noncommutative spaces have arisen in investigation
of brane configurations in string and M-theory. Since the
one-particle sector of field theories leads to QM, a study of this
topic has attracted much interest. For single particle QM, the
corresponding Heisenberg algebra is needed. In addition to
(\ref{djo-noncom}) and (\ref{djo-com}) one chooses
\begin{equation}
\label{djo-mom} [p^i,p^j]=i\Phi^{ij}.
\end{equation}
There are a lot of possibilities in choosing $\theta$ and
$\Phi$.\cite{djo-wess} Although one can take $\theta^{ij}$ and
$\Phi^{ij}$ to be antisymmetric nonconstant tensors (matrices),
often the simplest nontrivial case is considered:
$\theta^{ij}=const$ and $\Phi^{ij}=0$. Another realization of
noncommutativity is possible by $q$-deformation of a space, for
example, {\it Manin plane} $xy=qyx$ and $q$-deformed "classical"
phase space $px=qpx$. This approach leads to a latticelike
(discrete) structure of space-time.\cite{djo-wess1}

A field $\Psi(x)$ as a function of the noncommuting coordinates $x$
can be used as Schr\"odinger wave function obeying the free field
equation. Other realization, based on star product ($v\ast \Psi$)
instead of standard multiplication ($v\cdot \Psi$) of a potential
and wave function have been considered in corresponding
Schr\"odinger equation as well (i.e. see
Ref.~\refcite{djo-mezincesku}).

The passage from one level of physical theory to a more refined one,
using what mathematicians call deformation theory, is nothing
extraordinary new. In a similar way, there is an old idea that QM is
some kind of deformed classical mechanics. For a review see
Ref.~\refcite{djo-sternheimer}. In fact, deformation quantization is
closely related to Weyl quantization, briefly sketched out in the
previous section.

One direction of the investigation led to Moyal bracket and Moyal
(star) product,\cite{djo-waldmann} widely used now in noncommutative
QM
\begin{equation}
f\ast_m g=\chi_\infty
\left(
-\frac{h}{8\pi^2}P
\right)
(f,g)=
fg+\sum_{r=1}^{\infty}
\left(
\frac{ih}{4\pi}
\right)^r P^r(f,g).
\end{equation}
Several integral formulas have been introduced for the star product
and a (formal) parameter of deformation is finally related to some
form of Planck constant $h$. Quantization can be taken as a
deformation of the standard associative and commutative product, now
called a star product, of classical observables driven by the
Poisson bracket $P$. By the intuition, classical mechanics is
understood as the limit of QM when $h\rightarrow 0$.

Some connections between  $p$-adic analysis and quantum deformations
has been noticed. It has been observed that the Haar measure on
$SU_q(2)$ coincides with the Haar measure on the field of $p$-adic
numbers $\Bbb Q_p$ if $q=1/p$.\cite{djo-av} There is a potential
such that the spectrum of the $p$-adic Schr\"odinger-like
(diffusion) equation\cite{djo-vladi}
\begin{equation}
\label{djo-povisilica}
D\psi (x)+V(|x|_p)\psi (x)=E\psi (x)
\end{equation}
is the same as in the case of $q$-deformed oscillator found by
Biedenharn\cite{djo-b} for $q=1/p$. For more details see
Ref.~\refcite{djo-av}.

In a development of the representation theory for the star product
algebras in deformation quantization some non-Archimedean behavior
is noted\cite{djo-waldmann}. We find this relation very intriguing
and discuss it in more details. Recall that a star product $\ast $,
on a Poisson manifold (M,$\pi$) is a (formal) associative $\Bbb
C[[\lambda]]$ - bilinear product for $C^{\infty}(M)[[\lambda]]$
\begin{equation}
f\ast g = \sum_{r=0}^{\infty} \lambda^r C_r(f,g),
\end{equation}
with bidifferential operators $C_r$. $\Bbb C[[\lambda]]$ is an
algebra of {\it formal} (in a formal parameter $\lambda$) series
$\sum_{r=0}^{\infty} \lambda^r C_r, \ C_r\in \Bbb C$ (a convergence
of this series is still not under a consideration).
$C^{\infty}(M)[[\lambda]]$ is the space of formal series of smooth
functions ($x\in M,\ f_r(x)\in C^{\infty}(M)$), for a fixed but
variable $x$. With interpretation $\lambda \leftrightarrow \hbar$ we
identify $C^{\infty}(M)[[\lambda]]$ with the algebra of observables
of the quantum system corresponding to $(M,\pi)$. Let $R$ be on an
ordered (commutative associative unital) ring
$R$.\cite{djo-waldmann} Let us note that a concept of ordered ring
is necessary if one wishes to define the positive functionals on a
$C^\ast$ algebra ($C=R(i)=\{a+bi,\ a,b\in R\}$). It is related to
Gelfand-Naimark's theorem on commutative spaces. By means of the
positive functionals on $C^\ast$ algebras we can reconstruct the
(points on) ``starting'' manifold, the one on which an algebra of
complex function will give the ``original''  $C^\ast$ algebra. This
is a motivation for generalization on noncommutative space. In this
case, the corresponding product of formal functions will be the
$\ast$ product. Now, $R[[\lambda]]$ will be an algebra of formal
series on the ordered ring $R$. Then, if $R$ is an ordered ring,
$R[[\lambda]]$ will be ordered in a canonical way, too, by the
definition
\begin{equation}
\sum_{r=r_0}^{\infty} \lambda^ra_r>0,\ \ if \ \ a_{r_0}>0.
\end{equation}
In other words, a formal series in $R[[\lambda]]$ will be a
``positive`` one if first nonzero coefficient (an element of ordered
ring $R$ is positive). \noindent It should be noted that the concept
of ordered ring fits naturally with formal power series and thus to
Gerstenhaber's deformation theory.\cite{djo-gerstenhaber} Then
$R[[\lambda]]$ will be {\it non-Archimedean}! For example, if we
take that $a_0=-1$, and $a_1=n$ ($n\in \Bbb N$, because for any
commutative ordered ring with unit, set of integers is embedded in
$R$, $\Bbb N\subset R$), all other coefficients can be zero, we have
$-1+n\lambda<0$ or $n\lambda<1$ for all $n\in \Bbb Z$. The
interpretation in formal theory is that the deformation parameter
$\lambda$ is ``very small'' compared to the other numbers in R. We
see that Waldmann's definition of the ordered algebra of formal
series on ordered ring $R$ immediately leads to the non-archimedean
``structure''.

This result is a good indication to use ultrametric spaces and
$p$-adics when physical deformation parameter is very small. We
would like to underline that Zelenov's ordering by means of map
(\ref{map}) $\Bbb Q_p$, i.e. on normed, ultrametric algebra in frame
of $p$-adic QM for $\lambda=1/p^2$ is related to the ordering of
formal series on ordered rings in the frame of deformation
quantization! This is a quite intriguing coincidence of a strong
potential for further investigation.

\section{The Adelic Moyal Product and Noncommutative QM}

The presented connections between noncommutative vs.
"nonarchimedean" QM suggest a need to formulate a quantum theory
that may connect as much as possible nonarchimedean and
noncommutative effects and structures. At the present level of
quantum theory on adeles, a formulation of noncommutative adelic QM
seems to be the most promising attempt.

A simple enough frame for that might be the representation of an
algebra of operators (\ref{djo-noncom}), (\ref{djo-com}) and
(\ref{djo-mom}). It could be done by linear transformations on the
corresponding simplectic structure and deformed and extended
bilinear product $B$. Correspondence between classical functions and
quantum operators would be provided by Weyl quantisation. An
equivalent formulation of noncommutative adelic QM by the triple
$(L_2(A_\theta), W_\theta(z), U_\theta(t))$, does not seem to have
principles obstacles. In this approach an adele of coordinates $x_A$
would be replaced by a series of noncommutative operators $\hat
x_A$, where adelic properties of corresponding eigenvalues are still
"preserved".

Now, we have to consider a $p$-adic and adelic generalization of the
Moyal product. Let us consider classical space with coordinates
$x^1,x^2$, $\cdots,x^D$. Let $f(x)$ be a classical function
$f(x)=f(x^1,x^2,\cdots,x^D)$. Then, with respect to the Fourier
transformations and the usual Weyl quantization, we have
\begin{equation}
\hat f(x)=\int_{\Bbb Q_\infty^D} dk \
\chi_\infty(-k \hat x) \tilde f(k)\equiv  f(\hat x).
\end{equation}
Let us now have two classical functions $f(x)$ and $g(x)$ and we are
interested in operator product $\hat f(x) \hat g(x)$. In the real case this
operator product is
\begin{equation}
\label{djo-rmoyal}
(\hat f \cdot \hat g)(x)=\int \int dk dk' \ \chi_\infty(-k\hat x)
\chi_\infty(-k' \hat x) \tilde f(k)\tilde g(k').
\end{equation}
Using the Baker-Campbell-Hausdorff formula, the relation
(\ref{djo-noncom}) and then the coordinate representation one finds
the Moyal product in the form
\begin{equation}
(f\ast g)(x)=\int_{\Bbb Q_p^D}\int_{\Bbb Q_p^D} dk dk' \ \chi_\upsilon\left ( -(k+k')x+\frac 1 2
k_ik'_j\theta^{ij}\right )\tilde f(k)\tilde g(k').
\end{equation}
Note that we already used our generalization from  $\Bbb Q_\infty$
to $\Bbb Q_\upsilon$. In the real case we  use $k_i\rightarrow
-(i/2\pi)(\partial/\partial x^i)$ and obtain the well known form
$(f\ast g)(x)=\chi_\infty\left(-\frac{\theta^{ij}}{2(2\pi)^2}
\frac{\partial}{\partial y^i}\frac{\partial}{\partial z^j}\right )
f(y)g(z)|_{y=z=x}$. In the $p$-adic case such a straightforward
generalization is not possible\cite{kijev,dubrovnik} (but some kind
of pseudodifferentiation could be useful). Thus, as the $p$-adic
Moyal product we take
\begin{equation}
(f \ast g)(x)=\int_{\Bbb Q_p^D}\int_{\Bbb Q_p^D} dk dk'
\ \chi_p(-(x^ik_i+x^jk'_j)+\frac{1}{2} k_ik'_j\theta^{ij})\tilde f(k)\tilde g(k').
\end{equation}
We can write down the adelic Moyal product of "classical" adelic
functions $f_A=(f_\infty,f_2,...,f_p,...)$,
$g_A=(g_\infty,g_2,...,g_p,...)$ on $\Bbb R\times \prod_{p\in S}
\Bbb Q_p$ $\times \prod_{p\not\in S}\Bbb Z_p$ space
\begin{eqnarray}
(f \ast g)(x)=\int_{\Bbb A^D}\int_{\Bbb A^D} dk dk' \
\chi(-(x^ik_i+x^jk'_j)+\frac{1}{2} k_ik'_j\theta^{ij})\tilde
f_A(k)\tilde g_A(k').
\end{eqnarray}
Taking into account (\ref{djo-elementary}), (\ref{djo-rmoyal}) and the property
of the Fourier transform of $\Omega$ function, one has
\begin{eqnarray}
(f \ast g)(x)=\chi_\infty\left(-\frac{\theta^{ij}}{2(2\pi)^2}
\frac{\partial}{\partial y^i}\frac{\partial}{\partial z^j}\right ) f(y)g(z)|_{y=z=x}\nonumber \\
\times\prod_{p\in S} \int_{\Bbb Q_p^D}\int_{\Bbb Q_p^D} dk dk'
\ \chi_p(-(x^ik_i+x^jk'_j)+\frac{1}{2} k_ik'_j\theta^{ij})\tilde f_p(k)\tilde g_p(k')\nonumber \\
\times \prod_{p\notin S}\int_{\Bbb Z_p^D}\int_{\Bbb Z_p^D} dk dk'
\ \chi_p(-(x^ik_i+x^jk'_j)+\frac{1}{2} k_ik'_j\theta^{ij}).
\end{eqnarray}
It can be shown that if for all $p$, $\varphi(x)=\Omega(x)$, the
adelic Moyal product becomes real one, with some condition imposed
on $\theta$.

\section{Noncommutativity in Quantum Cosmology}

The words "quantum" and "cosmology" do appear to some physicists to
be inherently incompatible. We usually think of cosmology in terms
of the very large structure of the universe and of quantum phenomena
in terms of the very small. However, since gravity is the dominating
interaction on cosmic scales, a quantum theory of gravity is needed
as a formal prerequisite for quantum cosmology. Most work in quantum
cosmology is based on the Wheeler-DeWitt equation of quantum
geometrodynamics. In quantum mechanics and quantum field theory,
path integrals provide a convenient tool for a wide range of
applications. In quantum gravity, a path-integral formulation would
have to employ a sum over all four-metrics for a given topology.

In this paper we consider (4+D) Kaluza-Klein models, or more
precisely a particular, simpler model, the so called (4+1) ``empty``
model.

We start with the metric considered in
Refs.~\refcite{wudka,darabi1,darabi2,16blacksee}, in which spacetime
is of a Friedman-Robertson-Walker type, having a compactified space
which is assumed to be the circle $S^1$. We adopt the chart
$\{t,r^i,\rho\}$ with $t$, $r^i$ and $\rho$ denoting the time, the
space coordinates and the compactified space coordinate,
respectively
\begin{equation}
\label{metric} ds^2=-N^2dt^2+R^2(t)\frac{dr^idr^i}{(1+\frac{\kappa
r^2}{4})^2} +a^2d\rho^2,
\end{equation}
where $\kappa=0,\pm1$ and $R(t)$, $a(t)$ are the scale factors of
the universe and compact dimension, respectively. The integration of
the Einstein-Hilbert action for such an empty (4+1) dimensional
Kaluza-Klein universe with cosmological constant $\Lambda$
\begin{equation}
\label{dejstvo} S=\int\sqrt{-g}(\tilde R - \Lambda)dtd^3rd\rho,
\end{equation}
($\tilde R$ is a curvature scalar corresponding to metric
(\ref{metric})) over spatial dimensions gives an effective
Lagrangian in the minisuperspace $(R,a)$

\begin{equation}
L=\frac{1}{2N}Ra\dot R^2+\frac{1}{2N}R^2\dot R\dot a
-\frac{1}{2}N\kappa Ra + \frac{1}{6}N\Lambda R^3a.
\end{equation}

\subsection{Commutative model over real space}

By defining $\omega^2=-\frac{2\Lambda}{3}$ ($\Lambda<0$) and
changing variables as
\begin{equation}
u=\frac{1}{\sqrt{8}}[R^2+Ra-\frac{3\kappa}{\Lambda}], \ \
v=\frac{1}{\sqrt{8}}[R^2-Ra-\frac{3\kappa}{\Lambda}],
\end{equation}
in the "new" minisuperspace $(u,v)$, Lagrangian takes the form
\begin{equation}
L=\frac{1}{2N}(\dot u^2-N^2\omega^2u^2) -\frac{1}{2N}(\dot
v^2-N^2\omega^2v^2),
\end{equation}
which describes an isotropic oscillator-ghost-oscillator system.
Corresponding equations of motion and solutions are
\begin{equation}
\label{eom} \ddot u + N^2\omega^2 u=0,\ \ \ \ddot v +N^2\omega^2
v=0,
\end{equation}
 \begin{equation}
 u(t)=A\cos N\omega t+B\sin N\omega t , \ \ \ v(t)=C\cos N\omega
t+D\sin N\omega t.
 \end{equation}
Classical action is $$ \bar S(u'',v'',N;u',v',0) $$
\begin{equation}
=\frac{1}{2}\omega \left[ (u''^2+u'^2-v''^2-v'^2)\cot N\omega +
(v'v''-u'u'')\frac{2}{\sin N\omega} \right],
\end{equation}
\noindent while the propagator is
\begin{equation}
\label{prop} {\cal K}(u'',v'',N;u',v',0)=\frac{1}{i}
\sqrt{\frac{\omega^2}{\sin N\omega}} \times\exp \left( 2\pi i\bar
S(u'',v'',N;u',v',0) \right).
\end{equation}
To obtain the energy eigenstates and eigen vectors, we need to
recast the propagator (\ref{prop}) in a form that permits a direct
comparison with the spectral representation for the Feynman
propagator given by
\begin{equation}
{\cal
K}(u'',v'',N;u',v')=\Theta(N)\sum_{l}\Phi_l^{(m_1,m_2)}(u'',v'')
\Phi_l^{*(m_1,m_2)}(u',v')e^{-i\frac{NE_{n,m}}{\hbar}}.
\end{equation}
Corresponding Wheeler-Dewitt equation for this model in the
minisuperspace $(u,v)$ is ($N=1$)
\begin{equation}
\left(\frac{\partial^2}{\partial u^2}- \frac{\partial^2}{\partial
v^2}-\omega^2u^2+\omega^2v^2\right)\Psi(u,v)=0.
\end{equation}
It has oscillator-ghost-oscillator solutions belonging to the
Hilbert space ${\cal H}^{(m_1,m_2)}({\cal L}^2)$ as
\begin{equation}
\Psi^{(m_1,m_2)}(u,v)=\sum_{l=0}^\infty
c_l\Phi_l^{(m_1,m_2)}(u,v),
\end{equation}
with $m_1,m_2\ge0$ and $c_l\in C$. For more details about basis
solutions $\Phi_l^{(m_1,m_2)}(u,v)$ see Ref.~\refcite{darabi1}.

\subsection{Commutative model over $p$-adic space}

Let us consider this model on a $p$-adic space. The main relations
(\ref{metric}) and (\ref{dejstvo}) connected with the
(4+1)-dimensional model in the $p$-adic case are the same as in the
real case. Now, the effective $p$-adic valued Lagrangian in the
minisuperspace $(R,a)$ is
\begin{equation}
L=\frac{1}{2N}Ra\dot R^2+\frac{1}{2N}R^2\dot R\dot a
-\frac{1}{2}N\kappa Ra + \frac{1}{6}N\Lambda R^3a.
\end{equation}
Again, by defining $\omega^2=-\frac{2\Lambda}{3}$ ($\Lambda<0$) and
changing variables as
\begin{equation}
u=\frac{1}{\sqrt{8}}[R^2+Ra-\frac{3\kappa}{\Lambda}], \ \
v=\frac{1}{\sqrt{8}}[R^2-Ra-\frac{3\kappa}{\Lambda}],
\end{equation}
in the "new" minisuperspace $(u,v)$, Lagrangian takes the form
\begin{equation}
L=\frac{1}{2N}(\dot u^2-N^2\omega^2u^2) -\frac{1}{2N}(\dot
v^2-N^2\omega^2v^2).
\end{equation}
Corresponding equations of motion has formally the same form as
(\ref{eom}), but it's solutions have a different domain of
convergence\cite{djo-vladi,16blacksee}
\begin{equation}
 u(t)=A\cos N\omega t+B\sin N\omega t , \ \ \ v(t)=C\cos N\omega
t+D\sin N\omega t.
 \end{equation}
Classical action is $$ \bar S_p(u'',v'',N;u',v',0) $$
\begin{equation}
=\frac{1}{2}\omega \left[ (u''^2+u'^2-v''^2-v'^2)\cot N\omega +
(v'v''-u'u'')\frac{2}{\sin N\omega} \right].
\end{equation}
This action leads to the propagator\cite{79jrp}

\begin{equation}
\label{pad2} {\cal K}_p(v'', u'',N; v', u', 0) =
\frac{1}{|N|_p}\chi_p \left( \frac{\omega(u''^2 + u'^2 - v'^2 -
v''^2)}{ 2 \tan N\omega} + \frac{\omega(v'v'' - u'u'')}{\sin
N\omega} \right).
\end{equation}
In the $p$-adic region of convergence of analytic functions $\sin x$
and $\tan x$, which is $G_p = \{x\in Q_p : |x|_p \leq |2p|_p\}$,
exist vacuum states $\Omega(|u|_p)\Omega(|v|_p)$,
$\Omega(p^\nu|u|_p)$ $\times\Omega(p^\mu|v|_p)$, $\nu,\mu= 1, 2, 3,
···$, and also

\begin{equation}
\label{pad3} \Psi_p(x,y)=\left\{
\begin{array}{rl}
\delta(p^\nu-|x|_p)\delta(p^\mu-|y|_p),& |N|_p\le p^{2\nu-2},|N|_p\le p^{2\mu-2}\\
\delta(2^\nu-|x|_2)\delta(2^\mu-|y|_2),& |N|_2\le 2^{2\nu-3},
|N|_2\le 2^{2\mu-3},
\end{array}
\right.
\end{equation}
where $\mu,\nu= 0,-1,-2,\cdots.$

Some $4(=3+1)$-dimensional quantum cosmological models, which in
$p$-adic sector looks like two decoupled harmonic oscillators, were
analyzed in details in
Refs.~\refcite{vol2,kijev,dubrovnik,16blacksee}.

\subsection{Noncommutative case}

In the previous section we assumed that in $(u,v)$ minisuperspace
held
\begin{equation}
[u,v]=0, \ \ [u,p_u]=[v,p_v]=i\hbar, \ \ [p_u,p_v]=0,
\end{equation}
(generalized momenta are $p_u=\frac{\dot u}{N}$ and $p_v=\frac{\dot
v}{N}$). In the noncommutative case we deal with the same
Lagrangian, but with algebra
\begin{equation}
[u,v]=i\theta, \ \ [u,p_u]=[v,p_v]=i\hbar, \ \ [p_u,p_v]=0.
\end{equation}
By the transformation
\begin{equation}
u=u-\frac{\theta}{2}p_v,\ \ v=v+\frac{\theta}{2}p_u,
\end{equation}
we can represent this model as a commuting but with the Lagrangian
\begin{equation}
L_\theta=\frac{\omega^2}{\omega^2_\theta} \left[ \frac{1}{2N} \left(
\dot u^2-N^2\omega^2_\theta u^2 \right) - \frac{1}{2N} \left( \dot
v^2-N^2\omega^2_\theta v^2 \right) + \frac{1}{2N}\omega^2_\theta
\theta(u\dot v+\dot u v) \right],
\end{equation}
where
$\omega^2_\theta=\frac{\omega^2}{1+\frac{\omega^2\theta^2}{4}}$.
The equations of motion and solutions are
\begin{equation}
\ddot u + N^2\omega^2_\theta u=0,\ \ \ \ddot v +N^2\omega^2_\theta
v=0,
\end{equation}
\begin{equation}
 u(t)=A\cos N\omega_\theta t+B\sin N\omega_\theta t ,
 \ \ \ v(t)=C\cos N\omega_\theta t+D\sin N\omega_\theta t.
 \end{equation}
From the classical action $$ \bar
S_\theta(u'',v'',N;u',v',0)=\frac{1}{2}\omega
\sqrt{1+\frac{\omega^2\theta^2}{4}} $$
\begin{equation}
\times\left[ (u''^2+u'^2-v''^2-v'^2)\cot N\omega_\theta -
(u'u''-v'v'')\frac{2}{\sin N\omega_\theta} +
\frac{\theta\omega_\theta}{N}(u''v''-u'v') \right],
\end{equation}
we first get
\begin{equation}
\left|
\begin{array}{cc}
-\frac{\partial^2\bar S}{\partial u'\partial u''} & -\frac{\partial^2\bar S}{\partial u'\partial v''} \\
-\frac{\partial^2\bar S}{\partial v'\partial u''} &
-\frac{\partial^2\bar S}{\partial v'\partial v''}
\end{array}
\right|^{1/2}=\sqrt{1+\frac{\omega^2\theta^2}{4}}\frac{\omega}{|\sin
N\omega_\theta|},
\end{equation}
and finally $$ {\cal K}_\theta(u'',v'',N;u',v',0)=\frac{1}{i}
\sqrt{1+\frac{\omega^2\theta^2}{4}}\sqrt{\frac{\omega^2}{\sin
N\omega_\theta}} $$
\begin{equation}
\label{thetaprop} \times\exp \left( 2\pi i\bar
S_\theta(u'',v'',N;u',v',0) \right).
\end{equation}
In commutative regime (obtained for the $\theta=0$) we
have\cite{fdp}
\begin{equation}
{\cal K}(u'',v'',N;u',v',0)=\frac{1}{i} \sqrt{\frac{\omega^2}{\sin
N\omega}} \times\exp \left( 2\pi i\bar S(u'',v'',N;u',v',0)
\right).
\end{equation}

This example and remarkable result shows that exploring connections
between standard, nonarchimedean and noncommutative theory,
including quantum cosmology, are very interesting. Applications of
noncommutativity in cosmology remain current at
present,\cite{zejak,misel} with our belief that noncommutative and
nonlocal tachyonic models of inflation deserve more attention in
future.

\section*{Acknowledgments}

Work on this paper is partially supported by the ICTP - SEENET-MTP
grant PRJ-09 within the framework of the SEENET-MTP Network and by
the Ministry of Education and Science Republic of Serbia under
Grant 176021. A part of the whole research program was cofinanced
by the DFG grant ``Noncommutative space-time structure -
Cooperation with Balkan Countries'' and the DAAD project fellow
``Nonarchimedean Analysis, $p$-Adic Quantum Mechanics and Quantum
Deformation``. I would like to thank P. Aschieri, I. Bakovi\' c,
B. Dragovi\' c, B. Jurco, L. M\"{o}ller, Lj. Ne\v si\' c and S.
Waldmann for many useful discussions on topics considered in this
paper. My honest thankfulness to J. Dimitrijevi\' c-Savi\' c for
her continual help in proofreading of many of my and SEENET-MTP
publications. A personal gratitude to my younger collaborators and
colleagues D. Dimitrijevi\' c, M. Milo\v sevi\' c and J.
Stankovi\' c for their technical help in preparation of the
manuscript and their outstanding service in the SEENET-MTP Network
mission.

\end{document}